\documentclass[a4paper,onecolumn,11pt]{quantumarticle}
\pdfoutput=1
\usepackage[utf8]{inputenc}
\usepackage[english]{babel}
\usepackage[T1]{fontenc}
\usepackage{amsmath}
\usepackage{hyperref}
\usepackage{amssymb}
\usepackage{tikz}
\usepackage{lipsum}

\usepackage{etoolbox}
\usepackage{mathtools}
\usepackage{dcolumn}
\usepackage{bm}
\usepackage{verbatim}
\usepackage[english]{babel}
\usepackage{graphicx}%
\usepackage{multirow}%
\usepackage{amsmath}%
\usepackage[title]{appendix}%
\usepackage{xcolor}%
\usepackage{textcomp}%
\usepackage{booktabs}%
\usepackage{listings}%
\usepackage{bbold}
\usepackage{amsthm}
\usepackage[capitalise]{cleveref}
\DeclareMathOperator{\tr}{tr}
\DeclareMathOperator{\sn}{sn}
\DeclareMathOperator{\nc}{nc}
\DeclareMathOperator{\Jsc}{sc}

\DeclareMathOperator{\cn}{cn}
\DeclareMathOperator{\dn}{dn}
\DeclareMathOperator{\dc}{dc}
\DeclareMathOperator{\arcsinh}{arcsinh}
\DeclareMathOperator{\sech}{sech}
\DeclareMathOperator{\arctanh}{arctanh}

\begin{document}
\title{Bootstrapping Fermionic $R$-operators with Decorated Yang--Baxter Equations}
\author{Zhao Zhang}
\email{zhao.zhang@fys.uio.no}
\affiliation{Department of Physics, University of Oslo, P.O. Box 1048 Blindern, N-0316 Oslo, Norway}
\orcid{0000-0002-9425-732X}
\maketitle
\begin{abstract}Integrating infinitesimal generators of local dynamics into finite-time evolution is generally harder than extracting a local Hamiltonian from a quantum gate. For integrable one-dimensional quantum many-body systems, a recent bootstrap constructs one-parameter $R$-operators directly from Hamiltonians satisfying the Reshetikhin condition. However, this condition excludes important models such as the Hubbard model, whose two-parameter $R$-operator was obtained by Shastry from a decorated Yang--Baxter equation obeyed by its free-fermion sector. Here, I extend the bootstrap to $R$-operators constrained by both the ordinary and decorated Yang--Baxter equations. Their joint expansion produces additional Hamiltonian-level necessary conditions and recursively determines candidate higher-order terms, subject to two-site locality at each order. I also derive compatibility and factorization conditions under which a Shastry-type superposition solves the $RLL$ relation after deformation by an onsite involution. The construction recovers the known two-parameter $R$-operators of the Hubbard and XYh models, whose full Yang--Baxter equations follow from established tetrahedral Zamolodchikov algebra relations. For a superconducting Hubbard deformation, the scalar equations are generically incompatible, placing it outside this framework. These results provide diagnostics and a systematic route to fermionic $R$-operators beyond the Reshetikhin condition. General sufficiency, convergence of the formal bootstrap, and sufficiency of the deformation conditions remain open.
\end{abstract}

\section{Introduction}
\label{sec:intro}

Integrable one-dimensional (1D) quantum many-body systems are known to be dual to two-dimensional (2D) exactly solvable statistical lattice models. This duality is most pronounced for vertex- or face-type models, whose Boltzmann weights are specified by the $R$-matrix, although for edge models including Ising and Potts models an $R$-matrix can also be formulated \cite{pokrovskyBashilov,BazhanovaStroganov,Miao2023integrablequantum,MartinsRmat,zhang2026yangbaxterequationchiralpotts}. The classical statistical lattice model can alternatively be interpreted as a brickwork quantum circuit, where the $R$-matrix becomes the matrix representation of an $R$-operator satisfying the operator form Yang--Baxter equation (YBE). Under an appropriate physical parametrization and normalization, an $R$-operator may encode a unitary quantum gate describing finite-time evolution generated by an integrable local Hamiltonian. Therefore, determining the $R$-operator corresponds to integrating the infinitesimal generator of quantum dynamics, and the existence of such an $R$-operator is a signature of integrability.

Like in the quantum-classical duality, this is much more difficult than deriving the integrable Hamiltonian from the $R$-operator. Nevertheless, this is not impossible, owing to the stringent constraint imposed by the YBE. If the $R$-operator depends only on one spectral parameter, in which case it is also referred to as `of difference form' or `relativistic', expanding the YBE results in an infinite number of consistency relations. The lowest-order one among them involves directly the local Hamiltonian, and is called the Reshetikhin condition \cite{Kulish:1982aa}. Kennedy devised a method to turn the Reshetikhin condition into a direct Hamiltonian-level integrability test \cite{Kennedy:1992aa}, which has recently been generalized to the higher-order conditions that enable a bootstrapping program for constructing the $R$-operator from an integrable Hamiltonian alone \cite{zhang2026bootstrappingrmatrix}. The Reshetikhin condition has subsequently been shown to imply all the higher-order conditions, and hence the one-parameter YBE \cite{sanatani2026simplenecessarysufficientcondition}, using an analogous theorem on conserved charges \cite{hokkyo2025integrabilitysingleconservationlaw}.

None of these results applies to the general YBEs satisfied by $R$-operators that depend on two spectral parameters. Although the Reshetikhin condition including its higher-order counterparts \cite{zhang2026bootstrappingrmatrix} and the boost operator that generate the first non-trivial conserved charge from the Hamiltonian \cite{PhysRevLett.86.5096, takizawa2002ladderoperatorsintegrableonedimensional} can both be generalized to the two-parameter case, they assume the knowledge of a full continuous family of Hamiltonians parameterized by one of the spectral parameters, resulting from differentiating the $R$-operator. Hence, they are of little use for either testing the integrability of an individual Hamiltonian, or constructing the two-parameter $R$-operator from it. 

In fact, two-parameter $R$-operators were historically constructed with a completely different method. This was first done for the Hubbard Hamiltonian \cite{PhysRevLett.56.1529,PhysRevLett.56.2453,ShastryDYBE}, a paradigmatic example that violates the Reshetikhin condition, and then the XYh Hamiltonian \cite{FermiR,RmatforHubbard}, which does satisfy the Reshetikhin condition. What they have in common is that the two-body terms in their Hamiltonians describe either one or two copies of free fermions. As a result, their R-operators satisfy in addition to the YBE, another relation that is the YBE decorated by the onsite term in their respective Hamiltonians \cite{ShastryDYBE}. Shastry discovered that the $R$-operator for the total Hamiltonian satisfying the two-parameter YBE can be constructed by leveraging a linear superposition of the two YBEs. However, the construction still seems largely an art within Shastry's framework, and it is not clear whether it can be generalized to generic two-parameter $R$-operators.

The purpose of the current paper is to make Shastry's construction slightly more transparent and systematic, such that it can be applied to a broader class of Hamiltonians. The starting point of my extension is to apply the bootstrap method to both the YBE and decorated YBE, which yields strong extra constraints that substantially simplify the bootstrap procedure, allowing higher-order coefficients in the $R$-operator expansion to be expressed explicitly in terms of lower-order ones. The additional constraint from the decorated YBE is equivalent to a time-reversal symmetry (TRS) of the $R$-operator \cite{RmatforHubbard,fusion}. This symmetry operator appears in the Hubbard Hamiltonian as the onsite interaction operator that drives an integrable deformation from free to interacting fermions. In the case of the XYh model, the onsite term corresponds to the magnetic field that breaks the TRS, which corresponds to the other paradigm of mass-generation, or gap-opening deformation. The idea of TRS breaking and time-reversal symmetric mass generation has appeared in similar models for topological superconductors \cite{PhysRevB.94.165142}.

The second result of this paper is a sequence of necessary compatibility tests for a Shastry-type construction of a two-parameter $R$-operator. The first test concerns the two-body part of the Hamiltonian: Simultaneous expansion of the ordinary and decorated YBEs produces a locality condition that is more restrictive than the Reshetikhin condition but more general than Maassarani’s free-fermion algebra \cite{SUNHubbard,fusion}. When this test is satisfied, the joint expansion recursively determines candidate Taylor coefficients of a one-parameter $R$-operator, provided that the coefficients remain supported on two neighboring sites. The resulting object should initially be understood as a formal power series; the present work does not establish convergence or prove that the lowest-order condition alone implies all higher-order decorated-YBE constraints. 

The second test concerns an onsite deformation and is derived within a specific Shastry-type superposition ansatz. It requires certain commutators and anticommutators involving the one-parameter $R$-operator and a local involution to span spectral-parameter-independent operator directions. A further factorization condition is then required for the scalar superposition coefficient to exist. These conditions ensure consistency of the restricted YBE obtained by fixing one spectral parameter, equivalently the corresponding $RLL$ relation. Satisfaction of the full two-parameter YBE requires additional tetrahedral Zamolodchikov algebra relations or an independent verification. This final step is known to hold in the Hubbard and XYh examples but is not proved here in general.

The rest of the paper is organized as follows. Section~\ref{sec:DYBE} introduces the decorated YBE and the free-fermionic integrability condition. This condition is then used to iteratively construct the $R$-operator of an integrable Hamiltonian. The section concludes with explicit examples of $R$-operators for small local Hilbert-space dimension, which will be used in subsequent sections. Section~\ref{sec:Hubbard} carefully reviews Shastry’s construction of the Hubbard $R$-operator, with a new interpretation in terms of a time-reversal-symmetric mass-generation mechanism. A further highlight of this section is the derivation of explicit closed-form expressions via a new, systematic algebraic method. Section~\ref{sec:XYh} repeats this program for the XYh model demonstrating the time-reversal breaking gap-opening paradigm, where the trigonometric functions in the $R$-operator are generalized to Jacobi elliptic functions. As a by-product, differentiating the $R$-operator yields non-Hermitian integrable deformations of the XYh model, including the transverse-field Ising chain as a special case. The two successful constructions are complemented in Sec.~\ref{sec:coupleXY} by a superconducting deformation of the Hubbard Hamiltonian for which the proposed superposition ansatz fails. Analyzing this obstruction exposes an additional compatibility condition required by the construction, without excluding the possibility of a different integrable structure. Finally, Sec.~\ref{sec:conclusion} concludes by suggesting a few possible directions for future work.

\subsection{Summary and logical status of the results}

\begin{itemize}
    \item {Necessary Hamiltonian-level conditions for decorated YBE}:
\begin{itemize}
  \item Assuming the existence of an analytic one-parameter $R$-operator satisfying both the ordinary and decorated YBEs, their third-order expansions imply Eq.~\eqref{eq:dReshetikhin}.
  \item 
  Eq.~\eqref{eq:test} provides a direct test eliminating the unknown operators in Eq. \eqref{eq:dReshetikhin}, which is conjectured to be sufficient for higher-order tests to hold for \eqref{eq:higherj}.
\end{itemize}
\item {Formal recursive construction}:
\begin{itemize}
  \item Subject to two-site locality at every order, Eqs.~\eqref{eq:unieven} and \eqref{eq:higherR} recursively determine candidate Taylor coefficients of a formal one-parameter $R$-operator.
  \item It is conjectured, but not proved here, that the lowest-order conditions imply the complete decorated YBE.
\end{itemize}
\item {Onsite-deformation conditions within the superposition ansatz}:
\begin{itemize}
  \item Eqs.~\eqref{eq:condition1} or equivalently \eqref{eq:condition2}, and subsequently Eq.~\eqref{eq:generalf} are necessary for the operator-valued $RLL$ relation associated with the proposed Shastry-type ansatz.
  \item These conditions do not by themselves imply the full three-parameter YBE. For the Hubbard and XYh examples, the full YBE follows from their tetrahedral Zamolodchikov algebra relations. A general characterization of the additional assumptions under which the $RLL$ relation would imply the full YBE remains open.
\end{itemize}
\item {Examples and a counterexample}:
\begin{itemize}
  \item The construction reproduces the established Hubbard and XYh $R$-operators without using Shastry's ansatz on the explicit form of the $L$-matrix, and gives direct expressions for their superposition coefficients.
  \item For the superconducting Hubbard deformation, the proposed superposition ansatz produces incompatible scalar equations. This establishes failure of the present ansatz for generic parameters, not non-integrability of the Hamiltonian by all possible constructions.
\end{itemize}
\end{itemize}

\section{Integrable quantum circuits of free fermions}
\label{sec:DYBE}

\subsection{YBEs and time reversal symmetry}

\begin{figure}[ht]
	\centering
	\includegraphics[width=\linewidth]{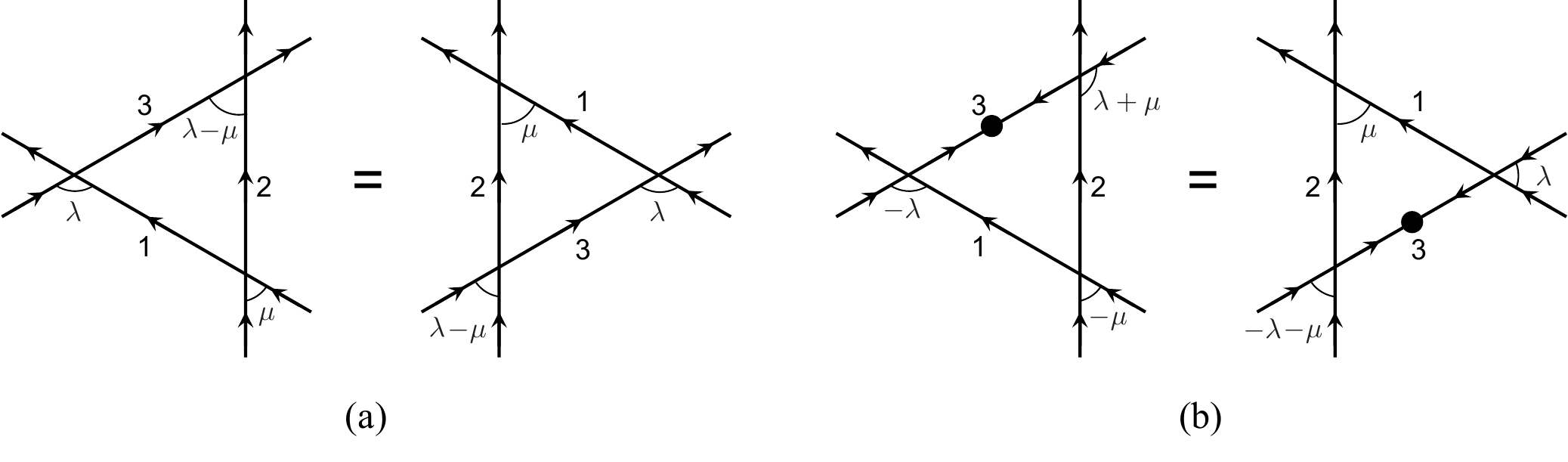}
	\caption{Diagrammatic representations of the YBE and decorated YBE. The positive time direction is upward, and a line with an arrow in the opposite direction can be interpreted as an anti-particle, with the sign of its rapidity reversed. (a) The YBE $R_{12}(\mu)R_{13}(\lambda)R_{23}(\lambda-\mu)=R_{23}(\lambda-\mu)R_{13}(\lambda)R_{12}(\mu)$ corresponding to the braided form \eqref{eq:YBE}. (b) The decorated YBE $R_{12}(\mu)R_{13}(\lambda)R_{23}(-\lambda-\mu)=R_{23}(\lambda+\mu)R_{13}(-\lambda)R_{12}(-\mu)$ or \eqref{eq:DYBE} in the braided form. The black dots in (b) denote the TRO $\Theta_3$ that appear in the original form of \eqref{eq:Shastry} $R_{12}(\mu)R_{13}(\lambda)\Theta_3R_{23}(\lambda+\mu)=R_{23}(\lambda+\mu)\Theta_3R_{13}(\lambda)R_{12}(\mu)$.}
	\label{fig:YBE}
\end{figure}

Shastry discovered in the course of constructing the two-dimensional solvable statistical model dual to the Hubbard Hamiltonian \cite{PhysRevLett.56.1529,PhysRevLett.56.2453} that the $R$-operator of the integrable XX chain satisfies in addition to the YBE (in the braided form) 
\begin{equation}
    \check{R}_{12}(\mu)\check{R}_{23}(\lambda)\check{R}_{12}(\lambda-\mu)=\check{R}_{23}(\lambda-\mu)\check{R}_{12}(\lambda)\check{R}_{23}(\mu)
\label{eq:YBE}
\end{equation}
a decorated YBE \cite{ShastryDYBE}
\begin{equation}
    \check{R}_{12}(\mu)\check{R}_{23}(\lambda)\Theta_2\check{R}_{12}(\lambda+\mu)=\check{R}_{23}(\lambda+\mu)\Theta_2\check{R}_{12}(\lambda)\check{R}_{23}(\mu),
\label{eq:Shastry}
\end{equation}where the decorating involution operator $\Theta$ satisfies $\Theta_j^2=1$ and implements $\lambda \mapsto -\lambda$ by conjugation
\begin{equation}
    \Theta_1 \check{R}_{12}(\lambda)\Theta_1=\Theta_{2} \check{R}_{12}(\lambda)\Theta_{2}=\check{R}_{12}(-\lambda). \label{eq:conjugation}
\end{equation}If the spectral parameter $\lambda$ is either interpreted as the rapidity or imaginary-time variable, $\Theta_j$ can be considered the time reversal operator (TRO) acting on site $j$. In the case of the free fermion chain, the TRO is naturally given by $\Theta_j=2n_j-1=2c_j^\dagger c_j-1$. Using the TRS, Shastry's decorated YBE can be equivalently written as \cite{RmatforHubbard,fusion}
\begin{equation}
    \check{R}_{12}(\mu)\check{R}_{23}(\lambda)\check{R}_{12}(-\lambda-\mu)=\check{R}_{23}(\lambda+\mu)\check{R}_{12}(-\lambda)\check{R}_{23}(-\mu).
\label{eq:DYBE}
\end{equation}So, \eqref{eq:YBE} and \eqref{eq:DYBE} can be considered as the two independent constraints. Since our first goal is to obtain an integrability condition on the local Hamiltonian without prior knowledge of the decorating involution, we will work with \eqref{eq:YBE} and \eqref{eq:DYBE}. Like the YBE, which has a graphical notation as shown in Fig.~\ref{fig:YBE} (a), the decorated YBE may also be diagrammatically represented by Fig.~\ref{fig:YBE} (b).

\subsection{Iterative solution of the \texorpdfstring{$R$}{R}-operator}

In the following, $\check{R}_{12}(\lambda)$ is assumed to be analytic in a neighborhood of $\lambda=0$, or alternatively to be treated as a formal power series
\begin{equation}
	\check{R}_{12}(\lambda)=\sum_{n=0}^\infty \frac{\lambda^n}{n!} \check{R}_{12}^{(n)},
	\label{eq:Rmat}
\end{equation}imposing $\check{R}_{12}^{(0)}=1$, and $\check{R}_{12}^{(1)}=h_{12}+e$, where $e$ is a constant shift of the energy density. No convergence theorem for the recursively obtained series is claimed. Accordingly, the output of the bootstrap will be called a formal candidate $R$-operator until a closed-form analytic realization or an independent YBE verification is supplied. 

The purpose of including the constant shift $e$ is to allow the $R$-operator to satisfy simultaneously the regularity condition $\check{R}_{12}(0)=1$ and the unitarity condition \begin{equation}
	\check{R}_{12}(\lambda)\check{R}_{12}(-\lambda)=1
	\label{eq:unitarity}
\end{equation}in an appropriate normalization. Eq.~\eqref{eq:unitarity} constrains even-order terms in the expansion \eqref{eq:Rmat} to be
\begin{equation}
	\check{R}_{12}^{(2m)}=\frac{1}{2}\sum_{k=1}^{2m-1}(-1)^{k-1}\binom{2m}{k}\check{R}_{12}^{(k)}\check{R}_{12}^{(2m-k)},
	\label{eq:unieven}
\end{equation}including the second order term $\check{R}_{12}^{(2)}=(h_{12}+e)^2$. 

The terms proportional to $\lambda \mu^2$ in \eqref{eq:YBE} and \eqref{eq:DYBE} give respectively
\begin{subequations}
  \begin{equation}
    [h_{12}+h_{23},[h_{12},h_{23}]] =\big(\check{R}_{23}^{(3)}-(h_{23}+e)^3\big)-\big(\check{R}_{12}^{(3)}-(h_{12}+e)^3\big),
	\label{eq:Reshetikhin}
  \end{equation}
  \begin{equation}
    \label{eq:ffic}
   [h_{12}-h_{23},[h_{12},h_{23}]] =\big(\check{R}_{23}^{(3)}-(h_{23}+e)^3\big)+\big(\check{R}_{12}^{(3)}-(h_{12}+e)^3\big).
  \end{equation}\label{eq:dReshetikhin}
\end{subequations}The former is the well-known Reshetikhin condition for integrable models with one-parameter $R$-operators \cite{Kulish:1982aa}, while the latter is an additional condition that is satisfied by integrable free fermions.  

For generic integrable Hamiltonians with one-parameter $R$-operators, a bootstrapping program has been proposed based on Kennedy's partial trace trick \cite{Kennedy:1992aa} to solve $\check{R}_{12}^{(3)}$ from \eqref{eq:Reshetikhin}, which ultimately constructs the full $\check{R}_{12}(\lambda)$ solely from the Hamiltonian \cite{zhang2026bootstrappingrmatrix}. However, in the case of free fermions, \eqref{eq:Reshetikhin} and \eqref{eq:ffic} can be readily combined to give
\begin{equation}
    \check{R}_{12}^{(3)}=[h_{23},[h_{23},h_{12}]]+(h_{12}+e)^3, \quad \text{and}\quad 
    \check{R}_{23}^{(3)}=[h_{12},[h_{12},h_{23}]]+(h_{23}+e)^3.\label{eq:R3}
\end{equation}The right-hand sides of Eq.~\eqref{eq:R3} are a priori supported on three neighboring sites. Therefore, Eq.~\eqref{eq:R3} can define a two-site Taylor coefficient only if the relevant double commutator acts trivially on the third (resp.~first) site. This support reduction is the actual necessary condition, since Eq.~\eqref{eq:dReshetikhin} involves the yet to be determined third-order coefficient. 

The higher-order operators in \eqref{eq:Rmat} can also be expressed in terms of recurrence relations. They can be extracted from the coefficient of the $\lambda \mu^{2m}$ term in \eqref{eq:YBE} and \eqref{eq:DYBE}, together giving 
\begin{equation}\begin{split}
      \check{R}^{(2m+1)}_{12}=&\sum_{k=0}^{2m}(-1)^k\binom{2m}{k}\check{R}^{(k)}_{23}\check{R}^{(1)}_{12}\check{R}^{(2m-k)}_{23}+\sum_{k=1}^{2m}(-1)^k\binom{2m}{k-1}\check{R}^{(k)}_{12}\check{R}^{(2m+1-k)}_{12},\\
	  \check{R}^{(2m+1)}_{23}=&\sum_{k=0}^{2m}(-1)^k\binom{2m}{k}\check{R}^{(k)}_{12}\check{R}^{(1)}_{23}\check{R}^{(2m-k)}_{12}+\sum_{k=1}^{2m}(-1)^k\binom{2m}{k-1}\check{R}^{(k)}_{23}\check{R}^{(2m+1-k)}_{23}.
  \end{split}  \label{eq:higherR}
\end{equation}Eq.~\eqref{eq:higherR} is a recursion for candidate Taylor coefficients rather than, by itself, an existence theorem for an $R$-operator. At every order, its three-site expression must reduce consistently to the corresponding two-site embedding. Even if this occurs to all orders, one must still distinguish a formal solution from a convergent analytic solution. The present paper verifies the recursion through closed-form realizations in the examples below but does not prove these properties for an arbitrary Hamiltonian satisfying Eq.~\eqref{eq:dReshetikhin}.

The TRS \eqref{eq:conjugation} requires that 
\begin{equation}
        \left\{\Theta_1, \check{R}^{(2m+1)}_{12}\right\}= \left\{\Theta_2, \check{R}^{(2m+1)}_{12} \right\}= \left[
        \Theta_1, \check{R}^{(2m)}_{12}\right]= \left[\Theta_2, \check{R}^{(2m)}_{12} \right]= 0.\label{eq:coeffinvol}
\end{equation}As a sanity check, for $m=1$ they are satisfied by the hopping Hamiltonian $h_{12}=c_1c^\dagger_2+c_2c^\dagger_1$, since $\{\Theta_j,c_j\}=\{\Theta_j,c^\dagger_j\}=0$, and $[\Theta_j,c_{j'}]=[\Theta_j,c^\dagger_{j'}]=0$ for $j'\ne j$. For a generic free-fermionic Hamiltonian, if TRO is known, this can be a way to determine the constant shift in $\check{R}^{(1)}_{12}$.

\subsection{Locality tests}\label{sec:test}

The recursion provides a sequence of locality obstructions that any analytic solution of the ordinary and decorated YBEs must satisfy. Define 
\begin{equation}
    j^{(1)}_{123}=[h_{12},[h_{12},h_{23}]]=\{h_{12}^2,h_{23}\}-2h_{12}h_{23}h_{12}.
\end{equation}Let the trace over site 1 be normalized by $\tr_1 I_1=1$, where $I_1$ is the identity operator acting on site 1, and define the conditional expectation onto operators acting trivially on site 1 by $\mathcal{E}_1(A_{123})=I_1\otimes \tr_1A_{123}$. An operator $A_{123}$ belongs to $I_1\otimes \mathrm{End}(V_2\otimes V_3)$ if and only if $\mathcal{E}_1(A_{123})=A_{123}$. Consequently, the support requirement for the second expression in Eq.~\eqref{eq:R3} is exactly
\begin{equation}
   \mathcal{E}_1 (j^{(1)}_{123})=j^{(1)}_{123}. \label{eq:test}
\end{equation}The corresponding condition for the other embedding is obtained by tracing over site 3. For a translation-invariant and reflection-symmetric density, the two tests are related; without these assumptions, both must be checked independently.

In principle, similar tests on 
\begin{equation}
    j^{(m)}=\sum_{k=0}^{2m}(-1)^k\binom{2m}{k}\check{R}^{(k)}_{12}\check{R}^{(1)}_{23}\check{R}^{(2m-k)}_{12} \label{eq:higherj}
\end{equation}for $m\ge 2$ might impose additional constraints. In light of the recent proof that the Reshetikhin condition \eqref{eq:Reshetikhin} implies the ordinary YBE \eqref{eq:YBE} \cite{sanatani2026simplenecessarysufficientcondition}, it is natural to ask whether Eq.~\eqref{eq:ffic} also implies the decorated YBE \eqref{eq:DYBE}. If this conjecture turns out to be true, Eq.~\eqref{eq:test} would be the only test needed for the bootstrap procedure describe above to carry through. Appendix \ref{sec:sufficiency} develops decorated analogues of the Sutherland equation, transfer matrix, and conserved charges that may be useful for a proof.

The support test in Eq.~\eqref{eq:test}, as written, is invariant under the scalar shift $h \mapsto h + eI$. The constant shift $e$ nevertheless affects the reconstruction of the Taylor coefficient through the polynomial terms $(h+eI)^n$, the imposed normalization, and the involution constraints in Eq.~\eqref{eq:coeffinvol}, and should therefore be treated as a normalization freedom in the full recursive construction, although not relevant for the lowest-order locality test. As long as there is a value of $e$ for which the higher-order counterparts of \eqref{eq:test} are satisfied, the Hamiltonian passes the locality hierarchy required by the present formal bootstrap.

A special case that fulfills \eqref{eq:test} is the so-called `free-fermion algebra' discovered by Maassarani \cite{fusion}
\begin{equation}
    h_{12}h_{23}h_{12}=0, \quad \text{and}\quad \{h_{12}^2,h_{23}\}=h_{23}. \label{eq:ffalgebra}
\end{equation}Likewise imposing $h_{23}h_{12}h_{23}=0$, $\{h_{23}^2,h_{12}\}=h_{12}$ and additionally $h_{12}^3= h_{12}$, (by assuming $e=0$) we have explicitly $\check{R}_{12}^{(3)}=2 h_{12}$, and by induction
\begin{equation}
   \check{R}^{(2m-1)}_{12}=\frac{ (-4)^{m}(1-4^{m})B_{2m}}{2m}h_{12}, \quad \text{and}\quad \check{R}^{(2m)}_{12}=(-1)^{m}E_{2m} h_{12}^2, \label{eq:induction}
\end{equation}where $B_k$ and $E_k$ are respectively the Bernoulli and Euler numbers that appear in the Taylor coefficients of tangent and secant functions. So in this case, the $R$-operator is given by
\begin{equation}
     \check{R}_{12}(\lambda)=1+(\sec(\lambda)-1)h^2_{12}+\tan(\lambda) h_{12}. \label{eq:Rmatfreefermion}
\end{equation}It has the properties of regularity $\check{R}_{12}(0)=1$ and unitarity $\check{R}_{12}(\lambda)\check{R}_{12}(-\lambda)=1$, and reproduces the free Hamiltonians by $\partial_\lambda \check{R}_{12}(\lambda)|_{\lambda=0}=h_{j,j+1}$. In addition, since all three terms commute with one another, it can be shown that 
\begin{equation}
    \check{R}_{12}(\lambda)\check{R}_{12}(\mu)=\check{R}_{12}(\mu)\check{R}_{12}(\lambda)=\check{R}_{12}(\arctan(\tan\lambda\sec\mu+\tan\mu\sec\lambda)).\label{eq:composition}
\end{equation}The composition of the spectral parameter hints a more transparent parametrization, by the duality transformation $\check{R}_{12}(\lambda)=\check{R}^*_{12}(\lambda^*)$, with
\begin{equation}
    \tan \lambda= \sinh\lambda^*, \quad \sec\lambda= \cosh \lambda^*, \quad \tanh\lambda= \sin\lambda^*, \quad \sech\lambda= \cos\lambda^*,
\label{eq:dualtransform}
\end{equation}In this notation, \eqref{eq:composition} becomes $\check{R}^*_{12}(\lambda^*)\check{R}^*_{12}(\mu^*)=\check{R}^*_{12}(\mu^*)\check{R}^*_{12}(\lambda^*)=\check{R}^*_{12}(\lambda^*+\mu^*)$.

Clearly, \eqref{eq:ffalgebra} is not a necessary condition for \eqref{eq:test} to hold. For instance, the Hamiltonian can be scaled by an overall factor $\alpha$, without affecting its free-fermionic nature. In this case, the algebra should be modified to $\{h_{12}^2,h_{23}\}=\alpha^2 h_{23}, \{h_{23}^2,h_{12}\}=\alpha^2 h_{12}$ and $h_{12}^3= \alpha^2 h_{12}$. They correspond to a rescaling of the spectral parameter in the expansion \eqref{eq:Rmat}.

\subsection{\texorpdfstring{$\mathfrak{su}(2)$}{su2} models}

A less trivial example of integrable fermions that is not described by \eqref{eq:ffalgebra} is the XY model, with local Hamiltonian $h_{12}=J_xX_1X_2+J_{y}Y_1Y_2$, where $X, Y$ are the Pauli-$x$ and -$y$ matrices. It satisfies \eqref{eq:test} by $[h_{12},[h_{12},h_{23}]]=4J_xJ_y(J_{y} X_2X_3+J_x Y_2Y_3)$. In this case, the conjugation can be chosen as the Pauli-$z$ operator $Z_j$, which satisfies $\{Z_j,h_{12}\}=0$ for $j=1,2$, $Z_1h_{12}=h_{12}Z_2$, and $Z_2h_{12}=h_{12}Z_1$. The JW transformed XY model, for $J_x=-2(1+\gamma)$, $J_y=-2(1-\gamma)$, is written in terms of fermion operators as
\begin{equation}
    H_\mathrm{XY}=-\sum_j \left(c_j^\dagger c_{j+1}+c_{j+1}^\dagger c_j+\gamma c_j^\dagger c_{j+1}^\dagger +\gamma   c_{j+1}c_j\right).
\end{equation}Its $R$-operator can be written in terms of Jacobi elliptic functions with modulus $|\gamma|$ as \cite{TetraZamalochikov,RmatforHubbard}
\begin{equation}
\begin{split}
     \check{R}_{12}(\mu;\gamma)=&(n_1 n_2+(1-n_1)(1-n_2))+\dc(\mu,|\gamma|)(n_1(1-n_2)+(1-n_1)n_2)\\ &-\sn(\mu,|\gamma|)\dc(\mu,|\gamma|)(c^\dagger_1c_2+c_2^\dagger c_1)-\gamma \sn(\mu,|\gamma|)(c^\dagger_1c^\dagger_2+c_2 c_1). \label{eq:RmatXY}
\end{split}
\end{equation}It satisfies the regularity condition $\check{R}_{12}(0;\gamma)=1$, and reproduces the Hamiltonian density $\partial_\mu \check{R}_{12}(\mu;\gamma)|_{\mu=0}=-(c^\dagger_1c_2+c_2^\dagger c_1+\gamma c^\dagger_1c^\dagger_2+\gamma c_2 c_1)$ as before. The scalar unitarity $\check{R}_{12}(\mu;\gamma)\check{R}_{12}(-\mu;\gamma)=\dn^2(\mu,|\gamma|)$ may be converted locally to the normalization in Eq.~\eqref{eq:unitarity} by division by $\dn(\mu,|\gamma|)$ away from its zeroes. Since $\check{R}_{21}(\mu;\gamma)=\check{R}_{12}(\mu;-\gamma)$, it is no longer symmetric. Analogous to \eqref{eq:composition}, it composes as 
\begin{equation}
\begin{split}
     \check{R}_{12}(\lambda)\check{R}_{12}(\mu)=\check{R}_{12}(\mu)\check{R}_{12}(\lambda)=&(1+\gamma^2\sn\lambda\sn\mu)h'^2_{12}+\dc\lambda\dc\mu(1+\sn\lambda\sn\mu)h^2_{12}\\ &+\dc\lambda\dc\mu(\sn\lambda+\sn\mu)h_{12}+\gamma(\sn\lambda+\sn\mu)h'_{12},
\end{split}   
\end{equation}where $h'_{12}= -c_j^\dagger c_{j+1}^\dagger - c_{j+1}c_j$. $\Theta_j=2c^\dagger_j c_j-1$ remains the TRO with the properties 
\begin{equation}
    \Theta_1\check{R}_{12}(\mu;\gamma)\Theta_{1}=\Theta_2\check{R}_{12}(\mu;\gamma)\Theta_{2}=\check{R}_{12}(-\mu;\gamma), \quad \text{and } \quad [\check{R}_{12}(\mu;\gamma), \Theta_1\Theta_2]=0.
\end{equation}

To obtain a closed-form general condition for integrable Hamiltonians satisfying the decorated YBE, we can parametrize a generic bilocal Hamiltonian that is symmetric about the two neighboring degrees of freedom using $\mathfrak{su}(n)$ generators $T^\alpha$ as
\begin{equation}
    h_{j,j+1}=a_{\alpha}T^\alpha_j T^\alpha_{j+1}+ b_\alpha (T_j^\alpha + T_{j+1}^\alpha), \label{eq:sunham}
\end{equation}where summation of repeated Greek indices from 1 to $n^2-1$ is implied. Following Ref.~\cite{zhang2026bootstrappingrmatrix}, the current operator can be expressed as
\begin{equation}
    [h_{12},[h_{12},h_{23}]]=u_{\alpha\beta\gamma}T_1^\alpha T_2^\beta T_3^\gamma+v_{\alpha\beta}T_1^\alpha T_2^\beta+w_{\alpha\beta}T_1^\alpha T_3^\beta+x_{\alpha\beta}T_2^\alpha T_3^\beta+y_\alpha T_1^\alpha+z_\alpha T_2^\alpha,
\end{equation}where the coefficients are functions of $a_\alpha, b_\alpha$ and the structure constants. The free fermion condition is thus equivalent to $u_{\alpha\beta\gamma}=v_{\alpha\beta}=w_{\alpha\beta}=y_\alpha =0$. In the case of $n=2$, the traceless Hermitian matrices $T_{1,2,3}$ become the Pauli matrices $X,Y,Z$, the structure constants reduce to the Levi-Civita symbols $\epsilon^{\alpha\beta}_\gamma$, and the coefficients are given by
\begin{equation}
    \begin{split}                         u_{\alpha\beta\gamma}=&a_{\gamma}b_\delta(a_{\alpha}\epsilon_\gamma^{\delta\varepsilon}\epsilon_\beta^{\alpha\varepsilon}+a_{\varepsilon}\epsilon_\beta^{\gamma\varepsilon}\epsilon_\alpha^{\delta\varepsilon}),\quad w_{\alpha\beta}=a_{\beta}a_{\gamma}a_{\delta}\epsilon_\alpha^{\gamma\delta}\epsilon_\beta^{\delta\gamma}, \quad y_\alpha= a_{\beta}a_{\gamma}b_\delta \epsilon_\alpha^{\beta\gamma}\epsilon_\delta^{\gamma\beta},\\
        v_{\alpha\beta}=&b_\gamma b_\delta (a_{\varepsilon}\epsilon_\beta^{\gamma\varepsilon}\epsilon_\alpha^{\delta\varepsilon}-a_{\alpha}\epsilon_\alpha^{\gamma\varepsilon}\epsilon_\beta^{\delta\varepsilon}), \quad x_{\alpha\beta}=a_{\beta}b_\gamma b_\delta \epsilon_\alpha^{\gamma\varepsilon}\epsilon_\beta^{\delta\varepsilon}, \quad z_\alpha= a_{\beta}a_{\beta}b_\gamma \epsilon_\alpha^{\beta\delta}\epsilon_\gamma^{\beta\delta}.
        \end{split}
\end{equation}

As already discussed in Ref.~\cite{zhang2026bootstrappingrmatrix}, there are four non-trivial scenarios for the Hamiltonian to be integrable in the first place. So we can find the decorated YBE integrable conditions case by case. Case one is the XYZ model ($b_1=b_2=b_3=0$). The remaining condition $w_{\alpha\beta}=0$ becomes $a_1a_2a_3=0$, which is satisfied by the XY model. Case two is the XYh model ($b_1=b_2=a_3=0$, $b_3\ne 0$). In this case, $v_{11}=y_{3}=0$ requires $a_1=a_2=0$. So it cannot satisfy decorated YBE unless it becomes the non-interacting model ($a_1=a_2=a_3=0$). Case three is the transverse field Ising model ($b_3=a_1=a_2=0$). Although the model is exactly solvable as free fermions, it cannot satisfy $u_{3\beta3}=0$ unless $b_1=b_2=0$. In fact, it will be shown in Sec.~\ref{sec:XYh} that, as a special case of the XYh model, its $R$-operator is a function of two spectral parameters. Lastly, the longitudinal field Ising model ($a_1=a_2=b_1=b_2=0$), which is classical or diagonal, trivially satisfies the decorated YBE, and is also not described by the Maassarani algebra \eqref{eq:ffalgebra}. So the moral of the story is that all free fermionic integrable models do not satisfy the decorated YBE, and the XY model discussed above is the only interesting $\mathfrak{su}(2)$ Hamiltonian within the parameterization \eqref{eq:sunham} that satisfies the decorated YBE but not the Maassarani algebra. A catch of the above classification is that the parameterization \eqref{eq:sunham} assumes the Hermiticity of the Hamiltonian. As will be shown in Appendix \ref{sec:NonHermitian}, there are non-Hermitian integrable fermions that satisfy the free fermion algebra \eqref{eq:ffalgebra}.

An important result for $n=2$ is that the YBE and decorated YBE are the only two linearly independent equations of the form $X_{23}\check{R}_{12}(\lambda)\check{R}_{23}(\mu)=\check{R}_{12}(\mu)\check{R}_{23}(\lambda)Y_{12}$, where $X_{12}=Y_{12}= \check{R}_{12}(\lambda-\mu)$ gives the YBE and $X_{12}=\check{R}_{12}(\lambda+\mu)\Theta_1$, $Y_{12}=\Theta_2\check{R}_{12}(\lambda+\mu)$ corresponds to the decorated YBE \cite{TetraZamalochikov,FermiR}. So a general solution of the intertwiner in the $RLL$ relation is of the form
\begin{equation}
\begin{split}
    X_{12}&=\alpha\check{R}_{12}(\lambda-\mu)+\beta\check{R}_{12}(\lambda+\mu)\Theta_1,\\
    Y_{12}&=\alpha\check{R}_{12}(\lambda-\mu)+\beta \Theta_2\check{R}_{12}(\lambda+\mu).
\end{split}
\end{equation}This can be used to construct the $R$-operator of the XYh Hamiltonian \cite{RmatforHubbard}, and will be the starting point of Shastry's solution of the $R$-operator for the Hubbard model reviewed in the next section.

\section{Time-reversal symmetric mass-generating deformation}\label{sec:Hubbard}

\subsection{\texorpdfstring{$R$-operator}{R-operator} of interacting fermions as a superposition of free fermions}

\begin{figure}[ht]
	\centering
	\includegraphics[width=0.7\linewidth]{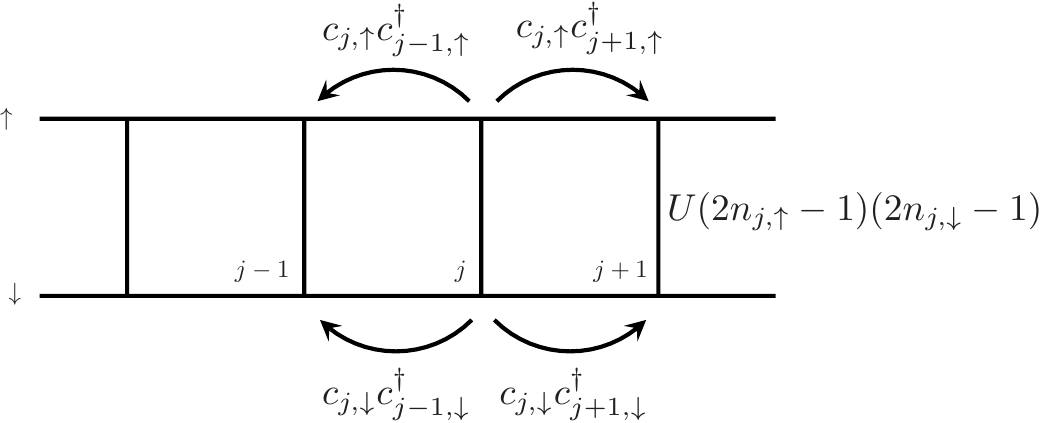}
	\caption{The Hubbard model viewed as two free fermion chains coupled by a diagonal interaction.}
	\label{fig:ladder}
\end{figure}
The 1D Hubbard model can be considered two copies of the free-fermion chain, one on each leg of a ladder, interacting with each other by the diagonal terms along the rungs, as depicted in Fig.~\ref{fig:ladder}. Its Hamiltonian is thus divided into three parts \begin{equation}
    H_\mathrm{Hubbard}(U)=\sum_j(h^\uparrow_{j,j+1}+h^\downarrow_{j,j+1}+Uh_{j}^\mathrm{int}),
\end{equation}with $h_{j,j+1}^\sigma=-c^\dagger_{j,\sigma}c_{j+1,\sigma}-c^\dagger_{j+1,\sigma}c_{j,\sigma}$ for $\sigma=\uparrow,\downarrow$, and $h_{j}^\mathrm{int}=\Theta^\uparrow_j\Theta^\downarrow_j$. The interaction part is expressed in terms of the TRO $\Theta_j^\uparrow=2n_{j,\uparrow}-1$, and $\Theta_j^\downarrow=2n_{j,\downarrow}-1$, where $n_{j,\sigma}=c^\dagger_{j,\sigma}c_{j,\sigma}$ are the fermion number operators. 

The $R$-operators for the free part of the Hamiltonian are given by 
\begin{equation}
\begin{split}
     \check{R}^\sigma_{12}(\lambda)=&1-P^\sigma_{12}+\sec(\lambda)P^\sigma_{12}+\tan(\lambda) h_{12}^\sigma, \label{eq:Rmatspinful}
\end{split}
\end{equation}with the projection operator $P^\sigma_{12}=(h_{12}^\sigma)^2=n_{1,\sigma}(1-n_{2,\sigma})+(1-n_{1,\sigma})n_{2,\sigma}$. Since they act non-trivially on subspaces with opposite spins, they mutually commute and their product $\check{R}^{\uparrow\downarrow}_{12}(\lambda)=\check{R}^\uparrow_{12}(\lambda)\check{R}^\downarrow_{12}(\lambda)$ satisfies simultaneously the YBE and decorated YBE. Motivated by the observation that a linear superposition of \eqref{eq:YBE} and \eqref{eq:Shastry} for $\check{R}^{\uparrow\downarrow}_{12}(\lambda)$ gives
\begin{equation}
\begin{split}
    \left( \check{R}^{\uparrow\downarrow}_{23}(\lambda-\mu)+f(\lambda,\mu)\check{R}^{\uparrow\downarrow}_{23}(\lambda+\mu)\Theta^\uparrow_2\Theta^\downarrow_2\right)\check{R}^{\uparrow\downarrow}_{12}(\lambda)\check{R}^{\uparrow\downarrow}_{23}(\mu)\\ =\check{R}^{\uparrow\downarrow}_{12}(\mu)\check{R}^{\uparrow\downarrow}_{23}(\lambda)\left( \check{R}^{\uparrow\downarrow}_{12}(\lambda-\mu)+f(\lambda,\mu) \Theta^\uparrow_2\Theta^\downarrow_2\check{R}^{\uparrow\downarrow}_{12}(\lambda+\mu)\right),
\end{split}\label{eq:linearcomb}
\end{equation}Shastry made the ingenious ansatz for the $R$-operator of the interacting Hubbard model that \cite{ShastryDYBE,RmatforHubbard}
\begin{equation}
    \check{R}_{12}(\lambda,\mu)=\check{R}^{\uparrow\downarrow}_{12}(\lambda-\mu)+f(\lambda,\mu)\check{R}^{\uparrow\downarrow}_{12}(\lambda+\mu)\Theta^\uparrow_1\Theta^\downarrow_1. \label{eq:ansatz}
\end{equation}The one- and two-parameter operators $\check{R}_{12}$ are distinguished by their arguments. 

\subsection{Elementary solution of the YBE for two-parameter \texorpdfstring{$R$-operators}{R-operators}}
\begin{figure}[ht]
	\centering
	\includegraphics[width=\linewidth]{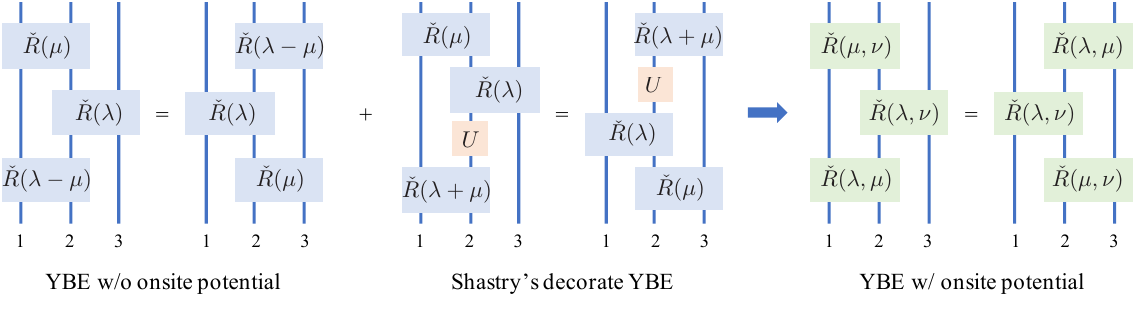}
	\caption{Shastry’s construction combines one-parameter $R$-operators for the two-body free-fermion sector, shown in blue, with the onsite interaction, shown in orange, to produce the two-parameter Hubbard $R$-operator, shown in green.}
	\label{fig:circuit}
\end{figure}
The strategy of Shastry's construction can be illustrated as shown in Fig.~\ref{fig:circuit} in the quantum circuit formulation of the YBEs. Plugging the ansatz \eqref{eq:ansatz} into the YBE
\begin{equation}
\check{R}_{23}(\lambda,\mu)\check{R}_{12}(\lambda, \nu)\check{R}_{23}(\mu,\nu) =\check{R}_{12}(\mu,\nu)\check{R}_{23}(\lambda,\nu)\check{R}_{12}(\lambda,\mu),
\label{eq:nonYBE}    
\end{equation}evaluated at $\nu=0$, and comparing with \eqref{eq:linearcomb}, we arrive at an equation for the unknown coefficient $f(\lambda,\mu)$,
\begin{equation}
\begin{split}
     \left( \check{R}^{\uparrow\downarrow}_{12}(\lambda-\mu)+f(\lambda,\mu)\Theta^\uparrow_2\Theta^\downarrow_2\check{R}^{\uparrow\downarrow}_{12}(\lambda+\mu)\right)\left(1+f(\lambda,0)\Theta^\uparrow_1\Theta^\downarrow_1\right)\left(1+f(\mu,0)\Theta^\uparrow_2\Theta^\downarrow_2\right)\\ 
   =\left(1+f(\mu,0)\Theta^\uparrow_1\Theta^\downarrow_1\right)\left(1+f(\lambda,0)\Theta^\uparrow_2\Theta^\downarrow_2\right)\left( \check{R}^{\uparrow\downarrow}_{12}(\lambda-\mu)+f(\lambda,\mu)\check{R}^{\uparrow\downarrow}_{12}(\lambda+\mu)\Theta^\uparrow_1\Theta^\downarrow_1\right).
\end{split}
\end{equation}Since $[\Theta^\uparrow_1\Theta^\downarrow_1\Theta^\uparrow_2\Theta^\downarrow_2,\check{R}^{\uparrow\downarrow}_{12}(\lambda)]=0$, the equation simplifies to
\begin{equation}
\begin{split}
    f(\lambda,\mu)\left[\check{R}^{\uparrow\downarrow}_{12}(\lambda+\mu)\left(\Theta^\uparrow_1\Theta^\downarrow_1+f(\lambda)f(\mu)\Theta^\uparrow_2\Theta^\downarrow_2\right)-\left(\Theta^\uparrow_2\Theta^\downarrow_2+f(\lambda)f(\mu)\Theta^\uparrow_1\Theta^\downarrow_1 \right)\check{R}^{\uparrow\downarrow}_{12}(\lambda+\mu)\right]\\
    =\check{R}^{\uparrow\downarrow}_{12}(\lambda-\mu)\left[f(\lambda)\Theta^\uparrow_1\Theta^\downarrow_1+f(\mu)\Theta^\uparrow_2\Theta^\downarrow_2\right]-\left[f(\lambda)\Theta^\uparrow_2\Theta^\downarrow_2+f(\mu)\Theta^\uparrow_1\Theta^\downarrow_1\right]\check{R}^{\uparrow\downarrow}_{12}(\lambda-\mu), \label{eq:equation}
\end{split}
\end{equation}with the shorthand notation $f(\lambda)=f(\lambda,0)$. Using 
\begin{equation}
    \begin{split}
        h_{12}^\sigma \Theta_1^\sigma&=\Theta^\sigma_2h_{12}^\sigma=-h_{12}^\sigma \Theta_2^\sigma=-\Theta^\sigma_1h_{12}^\sigma,\\
        \Theta_1^\sigma P^\sigma_{12}&=P^\sigma_{12}\Theta_1^\sigma=-\Theta_2^\sigma P^\sigma_{12}=-P^\sigma_{12}\Theta_2^\sigma=(\Theta_1^\sigma-\Theta_2^\sigma)/2,
    \end{split}\label{eq:chp}
\end{equation}we have 
\begin{equation}
    \begin{split}
        \left[ \check{R}^{\uparrow\downarrow}_{12}(\lambda), \Theta^\uparrow_1\Theta^\downarrow_1+\Theta^\uparrow_2\Theta^\downarrow_2\right]=&2\sec\lambda(\Theta^\uparrow_1\Theta^\downarrow_1-\Theta^\uparrow_2\Theta^\downarrow_2),\\
         \left\{ \check{R}^{\uparrow\downarrow}_{12}(\lambda), \Theta^\uparrow_1\Theta^\downarrow_1-\Theta^\uparrow_2\Theta^\downarrow_2\right\}=&2\sec\lambda\tan\lambda\left[h^\uparrow_{12}\Theta^\uparrow_1(\Theta^\downarrow_1-\Theta^\downarrow_2)+h^\downarrow_{12}(\Theta^\uparrow_1-\Theta^\uparrow_2)\Theta^\downarrow_2\right].
    \end{split}
\end{equation}Hence, \eqref{eq:equation} is reduced to 
\begin{equation}
    \begin{split}
        \sec(\lambda+\mu)f(\lambda,\mu)\left(1-f(\lambda)f(\mu)\right)&=\sec(\lambda-\mu)\left(f(\lambda)-f(\mu)\right),\\
         \tan(\lambda+\mu)\sec(\lambda+\mu)f(\lambda,\mu)\left(1+f(\lambda)f(\mu)\right)&=\tan(\lambda-\mu)\sec(\lambda-\mu)\left(f(\lambda)+f(\mu)\right).\label{eq:eqoff}
    \end{split}
\end{equation}The ratio between the two equations 
\begin{equation}
        \sin(2\mu) \frac{f(\lambda)}{1-f(\lambda)^2}=\sin(2\lambda) \frac{f(\mu)}{1-f(\mu)^2}
\end{equation}implies that 
\begin{equation}
    \frac{f(\lambda)}{1-f(\lambda)^2}\propto \sin(2\lambda).  \label{eq:propeq}
\end{equation}

Recall that the unitarity condition $\check{R}_{12}(\lambda,\mu)\check{R}_{12}(\mu,\lambda)\propto 1$ requires $f(\lambda,\mu)=-f(\mu,\lambda)$, which in turn tells us $f(\lambda,\lambda)=0$, consistent with the regularity condition $\check{R}_{12}(\lambda,\lambda)=1$. Since the derivative of $ \check{R}_{12}(\lambda,\mu)$ evaluated at $\lambda=\mu$ should reproduce the Hamiltonian, the constant of proportionality in \eqref{eq:propeq} is fixed by $\partial_\lambda f(\lambda,0)|_{\lambda=0}=U$, giving the unique solution
\begin{equation}
    f(\lambda,0)=\frac{\sqrt{1+U^2\sin^2(2\lambda)}-1}{U\sin(2\lambda)}=\tanh\left(\frac{\arcsinh(U\sin(2\lambda))}{2}\right).
\end{equation}Plugging this back into \eqref{eq:eqoff}, we have
\begin{equation}
\begin{split}
    f(\lambda,\mu)&=\frac{\cos(\lambda+\mu)}{\cos(\lambda-\mu)}\frac{f(\lambda,0)-f(\mu,0)}{1-f(\lambda,0)f(\mu,0)}\\
    &=\frac{\cos(\lambda+\mu)}{\cos(\lambda-\mu)}\tanh\left(\frac{\arcsinh\left(U\sin(2\lambda)\right)-\arcsinh\left(U\sin(2\mu)\right)}{2}\right)\\
    &=\frac{\cos(\lambda+\mu)}{\cos(\lambda-\mu)}\frac{U\left(\sin(2\lambda)-\sin(2\mu)\right)}{\sqrt{1+U^2\sin^2(2\lambda)}+\sqrt{1+U^2\sin^2(2\mu)}}\\
    &=\frac{\cot(\lambda+\mu)\left(\sqrt{1+U^2\sin^2(2\lambda)}-\sqrt{1+U^2\sin^2(2\mu)}\right)}{2U\cos^2(\lambda-\mu)}. \label{eq:f}
\end{split}
\end{equation}The new normalization is given by $\check{R}_{12}(\lambda,\mu)\check{R}_{12}(\mu,\lambda)=1-f(\lambda,\mu)^2$, which could be absorbed by a rescaling in the definition \eqref{eq:ansatz}.

Although the above derivation is largely overlapping with Shastry's original solution, by working directly with the YBE, it did not rely on his brilliant guess of the $L$-operator. The advantage is two-fold: First, the deductive approach makes it more promising to construct $R$-operators for other integrable interacting fermion models, which will be the topic of the next sections. Second, the deterministic approach establishes the uniqueness of the superposition coefficient within the ansatz \eqref{eq:ansatz}, normalization, gauge and twist transformations detailed in Section 12.2.5 of Ref.~\cite{Essler_Frahm_Korepin_2005}.

\subsection{The tetrahedral Zamolodchikov algebra}

A solution of \eqref{eq:nonYBE} at $\nu=0$, equivalently the associated $RLL$ relation, does not in general satisfy the full three-parameter YBE. For the Hubbard model, however, the solution in Eq.~\eqref{eq:ansatz} with the coefficient in Eq.~\eqref{eq:f} has been proved to satisfy Eq.~\eqref{eq:nonYBE} \cite{RmatforHubbard} through the tetrahedral Zamolodchikov algebra (TZA) \cite{TZA}
\begin{equation}
    \check{R}_{12}^{a,\sigma} (\mu,\nu) \check{R}_{23}^{b,\sigma} (\lambda,\nu)\check{R}_{12}^{c,\sigma}(\lambda,\mu)=\sum_{d,e,f=0,1}S_{def}^{abc}(\lambda,\mu,\nu)\check{R}_{23}^{f,\sigma} (\lambda,\mu)\check{R}_{12}^{e,\sigma} (\lambda,\nu)\check{R}_{23}^{d,\sigma}(\mu,\nu) ,
\end{equation}for $a,b,c=0,1$ and $\sigma=\uparrow,\downarrow$, where $\check{R}_{12}^{0,\sigma}(\lambda,\mu)=\check{R}^\sigma_{12}(\lambda+\mu)$, and $\check{R}^{1,\sigma}_{12}(\lambda,\mu)=\check{R}^\sigma_{12}(\lambda-\mu)\Theta_j$. Because of the high degree of symmetry among the 16 nonvanishing entries of the matrix $S$, only two out of the eight equations are independent. Four of them (with $(a+b+c)\mod2=0$) are equivalent to the YBE \eqref{eq:YBE} and decorated YBE \eqref{eq:DYBE} by the TRS \eqref{eq:conjugation}. The other four are equivalent to 
\begin{equation}
    \begin{split}
        &\check{R}^\sigma_{23}(\lambda+\mu)\Theta_2\check{R}^\sigma_{12}(\lambda+\nu)\Theta_1\check{R}^\sigma_{23}(\mu+\nu)\Theta_2\\
       = &S^{111}_{100}(\lambda,\mu,\nu)\check{R}^\sigma_{12} (\mu-\nu) \check{R}^\sigma_{23} (\lambda-\nu)\check{R}^\sigma_{12}(\lambda+\mu)\Theta_1\\
        &+S^{111}_{010}(\lambda,\mu,\nu)\check{R}^\sigma_{12} (\mu-\nu) \check{R}^\sigma_{23} (\lambda+\nu)\Theta_2\check{R}^\sigma_{12}(\lambda-\mu)\\
        &+S^{111}_{001}(\lambda,\mu,\nu)\check{R}^\sigma_{12} (\mu+\nu) \Theta_1\check{R}^\sigma_{23} (\lambda-\nu)\check{R}^\sigma_{12}(\lambda-\mu),
    \end{split}\label{eq:tetra2}
\end{equation}with
\begin{equation}
\begin{split}
    S^{111}_{100}(\lambda,\mu,\nu)=&-\tan(\mu+\nu)\cot(\lambda-\nu),\\
    S^{111}_{010}(\lambda,\mu,\nu)=&-\tan(\lambda+\mu)\tan(\mu+\nu),\\
    S^{111}_{001}(\lambda,\mu,\nu)=&\tan(\lambda+\mu)\cot(\lambda-\nu)
\end{split}
\end{equation}for the $R$-operators defined in \eqref{eq:Rmatspinful}. Equation \eqref{eq:equation} is the specialization of the relevant operator relations \eqref{eq:tetra2} to one vanishing spectral parameter. For the Hubbard solution, the complete TZA holds at arbitrary $\lambda,\mu,\nu$, and this additional three-parameter consistency ensures that Eq.~\eqref{eq:ansatz} satisfies the full YBE Eq.~\eqref{eq:nonYBE}. Interested readers are directed to Ref.~\cite{RmatforHubbard} for the complete proof (or Appendix 12.A of Ref.~\cite{Essler_Frahm_Korepin_2005} for a more pedagogical exposition).

\subsection{Non-Hermitian deformation of the Hubbard Hamiltonian}

If one differentiates the $R$-operator \eqref{eq:ansatz}, the outcome is a class of integrable Hamiltonians parametrized by one of the spectral parameters
\begin{widetext}
\begin{equation}
\begin{split}
        h_{12}(\mu)=\partial_\lambda \check{R}_{12}(\lambda,\mu)|_{\lambda=\mu}=h^\uparrow_{12}+h^\downarrow_{12}+\frac{U\cos^2(2\mu)}{\sqrt{1+U^2\sin^2(2\mu)}} \check{R}^{\uparrow\downarrow}_{12}(2\mu)\Theta^\uparrow_1\Theta^\downarrow_1.
\end{split}
\end{equation}
\end{widetext}It gives rise to a non-Hermitian Hamiltonian (except at $\mu=\frac{k\pi}{4}$, for $k\in \mathbb{Z}$) which, after telescopic cancellations and a constant shift, can be written as 
\begin{equation}
    H(\mu)=H_0+\frac{U}{\sqrt{1+U^2\sin^2(2\mu)}}\left[H_1(\mu)+H_2(\mu)-2N\right],\label{eq:nonHHubbard}
\end{equation}with $H_0=\sum_{j,\sigma}h^\sigma_{j,j+1}$, $H_1^\dagger(\mu)=H_1(\mu)$, $H_2^\dagger(\mu)=-H_2(\mu)$ and $N=\sum_{j,\sigma}n_j^\sigma$ commuting with the rest of the Hamiltonian. The Hermitian part is given by
\begin{equation}
    H_1(\mu)= 2\sum_{j}\left[\left(\cos^2(2\mu)+1\right)n_j^\uparrow n_j^\downarrow+\sin^2(2\mu)\left(Q_{j,j+1}^\uparrow Q_{j,j+1}^\downarrow-n_j^\uparrow n_{j+1}^\downarrow-n_j^\downarrow n_{j+1}^\uparrow\right)\right],
\end{equation}with the anti-Hermitian operator $Q_{j,j+1}^\sigma =c_{j,\sigma}^\dagger c_{j+1,\sigma}-c_{j+1,\sigma}^\dagger c_{j,\sigma}$ satisfying $[Q_{j,j+1}^\sigma,N]=0$, while the non-Hermitian part is given by
\begin{equation}
    \begin{split}
        H_2(\mu)=\sum_{j}\Bigg[&Q^\uparrow_{j,j+1}\left( \sin(4\mu)(n_j^\downarrow+n_{j+1}^\downarrow -1)+2\sin(2\mu)\left(n_j^\downarrow - n_{j+1}^\downarrow\right)\right) \\ +&Q^\downarrow_{j,j+1}\left( \sin(4\mu)(n_j^\uparrow+n_{j+1}^\uparrow -1)+2\sin(2\mu)\left(n_j^\uparrow - n_{j+1}^\uparrow\right)\right)\Bigg]. \label{eq:antiHHubbard}
    \end{split}
\end{equation}

\section{Time-reversal breaking mass-generating deformation} \label{sec:XYh}

Historically, the Hubbard model was the first integrable fermionic model to have its two-parameter $R$-operator constructed, rightfully so considering its central importance to condensed matter physics. But it is not the simplest model with such $R$-operators. In fact, it could have been done for spinless fermions with a chemical potential, or equivalently the XX model in a longitudinal magnetic field. Instead of repeating the simpler derivation than the previous section, I will work out a more general case, with anisotropic coupling in the $x$- and $y$-direction, which has been studied in Ref.~\cite{FermiR,RmatforHubbard}.

The JW-transformed Hamiltonian is given by 
\begin{equation}
    H_\mathrm{XYh}(\gamma)=-\sum_j (h_{j,j+1}+\gamma h'_{j,j+1})+U\Theta_j, \label{eq:XYhHam}
\end{equation}with $h_{j,j+1}=c^\dagger_{j}c_{j+1}+c^\dagger_{j+1}c_{j}$, $h'_{j,j+1}=c^\dagger_{j}c^\dagger_{j+1}+c_{j+1}c_{j}$, and $\Theta_j=2c_j^\dagger c_j-1$. Since the $R$-operator for $U=0$ is of the form \eqref{eq:RmatXY}, the general solution of the two-parameter YBE 
\begin{equation}
\check{R}_{23}(\lambda,\mu;\gamma)\check{R}_{12}(\lambda, \nu;\gamma)\check{R}_{23}(\mu,\nu;\gamma) =\check{R}_{12}(\mu,\nu;\gamma)\check{R}_{23}(\lambda,\nu;\gamma)\check{R}_{12}(\lambda,\mu;\gamma),
\label{eq:XYhnonYBE}    
\end{equation}
can be written as
\begin{equation}
    \check{R}_{12}(\lambda,\mu;\gamma)=\check{R}_{12}(\lambda-\mu;\gamma)+f(\lambda,\mu;\gamma)\check{R}_{12}(\lambda+\mu;\gamma)\Theta_1.\label{eq:XYhansatz}
\end{equation}Comparing \eqref{eq:XYhnonYBE} at $\nu=0$ with the linear combination of the YBE and decorated YBE satisfied by \eqref{eq:RmatXY}
\begin{equation}
    \begin{split}
        \left[ \check{R}_{23}(\lambda-\mu;\gamma)+f(\lambda,\mu;\gamma,U)\check{R}_{23}(\lambda+\mu;\gamma)\Theta_2\right]\check{R}_{12}(\lambda;\gamma)\check{R}_{23}(\mu;\gamma)\\ =\check{R}_{12}(\mu;\gamma)\check{R}_{23}(\lambda;\gamma)\left[ \check{R}_{12}(\lambda-\mu;\gamma)+f(\lambda,\mu;\gamma) \Theta_2\check{R}_{12}(\lambda+\mu;\gamma)\right],
    \end{split}
\end{equation}we get an equation on the coefficient $f(\lambda,\mu;\gamma,U)$
\begin{equation}
\begin{split}
     \left( \check{R}_{12}(\lambda-\mu;\gamma)+f(\lambda,\mu;\gamma)\Theta_2\check{R}_{12}(\lambda+\mu;\gamma)\right)\left(1+f(\lambda,0;\gamma)\Theta_1\right)\left(1+f(\mu,0;\gamma)\Theta_2\right)\\ 
   =\left(1+f(\mu,0;\gamma)\Theta_1\right)\left(1+f(\lambda,0;\gamma)\Theta_2\right)\left( \check{R}_{12}(\lambda-\mu;\gamma)+f(\lambda,\mu;\gamma)\check{R}_{12}(\lambda+\mu;\gamma)\Theta_1\right).
\end{split}\label{eq:XYhequation}
\end{equation}The operator equation is reduced to 
\begin{equation}
\begin{split}
    f(\lambda,\mu;\gamma)=&\frac{\dc(\lambda-\mu,|\gamma|)}{\dc(\lambda+\mu,|\gamma|)}\frac{f(\lambda,0;\gamma)-f(\mu,0;\gamma)}{1-f(\lambda,0;\gamma)f(\mu,0;\gamma)},\\
    f(\lambda,\mu;\gamma) =&\frac{\sn(\lambda-\mu,|\gamma|)}{\sn(\lambda+\mu,|\gamma|)}\frac{f(\lambda,0;\gamma)+f(\mu,0;\gamma)}{1+f(\lambda,0;\gamma)f(\mu,0;\gamma)}
\end{split}\label{eq:XYheqoff}
\end{equation}by the identities
\begin{equation}
    \begin{split}
         \check{R}_{12}(\mu;\gamma)\Theta_1-\Theta_2\check{R}_{12}(\mu;\gamma)=&\dc(\mu,|\gamma|)(\Theta_1-\Theta_2)+2\gamma \sn(\mu,|\gamma|)h'_{12}\Theta_1,\\
        \check{R}_{12}(\mu;\gamma)\Theta_2-\Theta_1\check{R}_{12}(\mu;\gamma)=&-\dc(\mu,|\gamma|)(\Theta_1-\Theta_2)+2\gamma \sn(\mu,|\gamma|)h'_{12}\Theta_1.
    \end{split}
\end{equation}After applying the addition rules of elliptic functions, \eqref{eq:XYheqoff} implies
\begin{equation}
    \Jsc(2\mu,|\gamma|)\frac{f(\lambda,0;\gamma)}{1-\left(f(\lambda,0;\gamma)\right)^2}=\Jsc(2\lambda,|\gamma|)\frac{f(\mu,0;\gamma)}{1-\left(f(\mu,0;\gamma)\right)^2},
\end{equation}which is solved by
\begin{equation}
    f(\lambda,0;\gamma)=\frac{\sqrt{1+U^2\Jsc^2(2\lambda,|\gamma|)}-1}{U\Jsc(2\lambda,|\gamma|)}=\tanh\left(\frac{\arcsinh(U\Jsc(2\lambda,|\gamma|))}{2}\right), 
\end{equation}such that $\partial_\lambda f(\lambda,0;\gamma,U)|_{\lambda=0}=U$ reproduces the correct Hamiltonian. Plugging the solution into \eqref{eq:XYheqoff} and repeating a derivation analogous to \eqref{eq:f}, we have found
\begin{equation}
\begin{split}
    f(\lambda,\mu;\gamma)=&\frac{\dc(\lambda-\mu,|\gamma|)}{\dc(\lambda+\mu,|\gamma|)}\tanh\left[\frac{\arcsinh(U\Jsc(2\lambda,|\gamma|))-\arcsinh(U\Jsc(2\mu,|\gamma|))}{2}\right]\\ 
    =&\frac{\dc(\lambda-\mu,|\gamma|)}{\dc(\lambda+\mu,|\gamma|)}\frac{\sqrt{1+U^2\Jsc^2(2\lambda,|\gamma|)}-\sqrt{1+U^2\Jsc^2(2\mu,|\gamma|)}}{U\left[\Jsc(2\lambda,|\gamma|)+\Jsc(2\mu,|\gamma|)\right]}. \label{eq:XYf}
\end{split}
\end{equation}According to Table \ref{tab:table1}, at $\gamma=1$, the elliptic functions degenerate to hyperbolic ones, and $f(\lambda,\mu;1)=\frac{\sqrt{1+U^2\sinh^2(2\lambda)}-\sqrt{1+U^2\sinh^2(2\mu)}}{U\left[\sinh(2\lambda)+\sinh(2\mu)\right]}$ gives the $R$-operator for the transverse field Ising model
\begin{equation}
    \check{R}_{12}(\lambda,\mu;1)=1-\tanh(\lambda-\mu)(h_{12}+ h'_{12})+f(\lambda,\mu;1)\left(1-\tanh(\lambda+\mu)(h_{12}+h'_{12})\right)\Theta_1.
\end{equation}

\begin{table}[ht]
\centering
\caption{\label{tab:table1}%
Special cases of Jacobi elliptic functions.
}\renewcommand{\arraystretch}{1}
\begin{tabular}{|c |c|c|c|c |c|}\cline{1-6}
$\gamma$ & $\sn$ & $\cn$ & $\mathrm{sc}$ & $\dn$ & $\dc$ \\ \cline{1-6}
0 & $\sin$ & $\cos$ & $\tan$ & $1$ & $\sec$  \\ \cline{1-6}
1 & $\tanh$  & $\sech$&$\sinh$ & $\sech$ & $1$ \\ 
 \cline{1-6}
\end{tabular}
\end{table}

From the $R$-operator we can obtain a class of integrable non-Hermitian Hamiltonians
\begin{equation}
        h_{12}(\mu;\gamma)=\partial_\lambda \check{R}_{12}(\lambda,\mu;\gamma)|_{\lambda=\mu}=-h_{12}-\gamma h'_{12}+\frac{U\nc(2\mu,|\gamma|)}{\sqrt{1+U^2\Jsc^2(2\mu,|\gamma|)}}\check{R}_{12}(2\mu;\gamma)\Theta_1.
\end{equation}After telescopic cancellation and a constant shift, the total Hamiltonian can be written as 
\begin{equation}
\begin{split}
     H(\mu;\gamma)=&H_\mathrm{XY}(\gamma)\\ &+\frac{U\nc(2\mu,|\gamma|)}{\sqrt{1+U^2\Jsc^2(2\mu,|\gamma|)}}\sum_j\left[\Theta_j+\sn(2\mu,|\gamma|)\left(\dc(2\mu,|\gamma|)Q_{j,j+1}+\gamma Q'_{j,j+1}\right)\right],
\end{split}
\end{equation}with the anti-Hermitian operators $Q_{j,j+1}=c^\dagger_jc_{j+1}-c^\dagger_{j+1} c_j$ and $Q'_{j,j+1}=c^\dagger_jc^\dagger_{j+1}-c_{j+1}c_j$.

The solution \eqref{eq:XYhansatz} with \eqref{eq:XYf} has been shown to satisfy the full YBE using the TZA \cite{TetraZamalochikov,FermiR}. In Ref.~\cite{FermiR}, the idea of introducing interaction to a single species of free fermion with the TRO was also applied to the Hubbard model, where an $R$-operator was obtained for the Hubbard model with nonzero chemical potential for both spins. Finally, as has been done for the Hubbard model \cite{PhysRevLett.86.5096}, a ladder operator $B=-\sum_j jh_{j,j+1}(\mu;\gamma)+\partial_\mu$ can be used to generate an infinite number of local conserved charges $I^{(n+1)}=[B,I^{(n)}]$ for $n\ge 2$, with $I^{(2)}=H(\mu;\gamma)$.

The construction above establishes a two-parameter representation of the XYh $R$-operator for the distribution of the onsite operator in Eq.~\eqref{eq:XYhHam}. A symmetric distribution of the onsite term in the XYh Hamiltonian actually satisfies the Reshetikhin condition. So a one-parameter solution exists from the bootstrap program \cite{zhang2026bootstrappingrmatrix}. The present derivation does not establish whether the two solutions are related by reparameterization, a local gauge transformation, or a more general twist. In particular, the existence of the two-parameter form obtained here should not be interpreted as proving that two-parameter dependence is essential. 

\section{Compatibility framework for onsite deformations using decorated YBEs}\label{sec:coupleXY}
In this section, I try to summarize the previous two examples into a general procedure for constructing two-parameter $R$-operators from one-parameter ones, first with a counterexample showing when it fails, and then proposing a multi-step compatibility test for the superposition ansatz.

\subsection{Superconducting Hubbard model from coupling XY chains}

After witnessing the success of the last two sections, it is natural to expect that the Hubbard model might be deformed with a superconducting pair-creation and annihilation term while remaining integrable. Similar Hamiltonians describing coupled Kitaev chains have been studied in Ref.~\cite{MattisNam,francica2019phasediagramquenchdynamics}. It should not be too surprising that Shastry's construction turns out not to apply to this generalization, as the decorated YBE and existence of a decorating involution are not sufficient for the present superposition construction. Nevertheless, since examining where the recipe fails is helpful for identifying further necessary compatibility conditions within the superposition ansatz, I will still carry out the detailed procedure for constructing an $R$-operator for the deformation. 

A superconducting Hubbard model is defined by the Hamiltonian 
\begin{equation}
    H(\gamma)=H_\mathrm{Hubbard}+\gamma\sum_j (h'^{\uparrow}_{j,j+1}+h'^{\downarrow}_{j,j+1}), \label{eq:superH}
\end{equation}with $h'^\sigma_{j,j+1}=-c^\dagger_{j,\sigma}c^\dagger_{j+1,\sigma}-c_{j+1,\sigma}c_{j,\sigma}$. The additional pair-creation/annihilation operators satisfy $(h'^\sigma_{j,j+1})^2=1-P_{j,j+1}^\sigma$. Using the $R$-operator for each leg \eqref{eq:RmatXY}
\begin{equation}
     \check{R}^\sigma_{12}(\mu;\gamma)=1-P_{12}^\sigma+\dc(\mu,\gamma)P_{12}^\sigma+\sn(\mu,\gamma)\dc(\mu,\gamma)h_{12}^\sigma+\gamma \sn(\mu,\gamma)h'^\sigma_{12},
\end{equation}we can make the same ansatz as before
\begin{equation}
    \check{R}_{12}(\lambda,\mu;\gamma)=\check{R}^{\uparrow\downarrow}_{12}(\lambda-\mu;\gamma)+f(\lambda,\mu;\gamma)\check{R}^{\uparrow\downarrow}_{12}(\lambda+\mu;\gamma)\Theta^\uparrow_1\Theta^\downarrow_1,\label{eq:newansatz}
\end{equation}with $\check{R}^{\uparrow\downarrow}_{12}(\mu;\gamma)=\check{R}^{\uparrow}_{12}(\mu;\gamma)\check{R}^{\downarrow}_{12}(\mu;\gamma)$. The YBE of the $R$-operator for the ladder system
\begin{equation}
\check{R}_{23}(\lambda,\mu;\gamma)\check{R}_{12}(\lambda, \nu;\gamma)\check{R}_{23}(\mu,\nu;\gamma) =\check{R}_{12}(\mu,\nu;\gamma)\check{R}_{23}(\lambda,\nu;\gamma)\check{R}_{12}(\lambda,\mu;\gamma),
\label{eq:newnonYBE}    
\end{equation}at $\nu=0$, together with the linear combination of the YBE and decorated YBE satisfied by the composite $R$-operator of the two legs of the ladder\begin{widetext}
\begin{equation}
    \begin{split}
        \left[ \check{R}^{\uparrow\downarrow}_{23}(\lambda-\mu;\gamma)+f(\lambda,\mu;\gamma)\check{R}^{\uparrow\downarrow}_{23}(\lambda+\mu;\gamma)\Theta^\uparrow_2\Theta^\downarrow_2\right]\check{R}^{\uparrow\downarrow}_{12}(\lambda;\gamma)\check{R}^{\uparrow\downarrow}_{23}(\mu;\gamma)\\ =\check{R}^{\uparrow\downarrow}_{12}(\mu;\gamma)\check{R}^{\uparrow\downarrow}_{23}(\lambda;\gamma)\left[ \check{R}^{\uparrow\downarrow}_{12}(\lambda-\mu;\gamma)+f(\lambda,\mu;\gamma) \Theta^\uparrow_2\Theta^\downarrow_2\check{R}^{\uparrow\downarrow}_{12}(\lambda+\mu;\gamma)\right],
    \end{split}
\end{equation}\end{widetext}gives with shorthand notation $f(\lambda;\gamma)=f(\lambda,0;\gamma)$,
\begin{equation}
\begin{split}
     &\left( \check{R}^{\uparrow\downarrow}_{12}(\lambda-\mu;\gamma)+f(\lambda,\mu;\gamma)\Theta^\uparrow_2\Theta^\downarrow_2\check{R}^{\uparrow\downarrow}_{12}(\lambda+\mu;\gamma)\right)\left(1+f(\lambda;\gamma)\Theta^\uparrow_1\Theta^\downarrow_1\right)\left(1+f(\mu;\gamma)\Theta^\uparrow_2\Theta^\downarrow_2\right)\\ 
   =&\left(1+f(\mu;\gamma)\Theta^\uparrow_1\Theta^\downarrow_1\right)\left(1+f(\lambda;\gamma)\Theta^\uparrow_2\Theta^\downarrow_2\right)\left( \check{R}^{\uparrow\downarrow}_{12}(\lambda-\mu;\gamma)+f(\lambda,\mu;\gamma)\check{R}^{\uparrow\downarrow}_{12}(\lambda+\mu;\gamma)\Theta^\uparrow_1\Theta^\downarrow_1\right).
\end{split}\label{eq:newequation}
\end{equation}Using \eqref{eq:chp} along with $h'^\sigma_{12} \Theta_1^\sigma=h'^\sigma_{12} \Theta_2^\sigma=-\Theta^\sigma_1 h'^\sigma_{12}=-\Theta^\sigma_2h'^\sigma_{12}$, we have 
\begin{equation}
    \begin{split}
        \left[ \check{R}^{\uparrow\downarrow}_{12}(\mu;\gamma), \Theta^\uparrow_1\Theta^\downarrow_1+\Theta^\uparrow_2\Theta^\downarrow_2\right]=&4\gamma  \sn^2(\mu,|\gamma|)\dc(\mu,|\gamma|) \left(h^\uparrow_{12}\Theta^\uparrow_1 h'^\downarrow_{12}\Theta^\downarrow_1+h'^\uparrow_{12}\Theta^\uparrow_1 h^\downarrow_{12}\Theta^\downarrow_1\right)\\
         &+2\dc(\mu,|\gamma|)(\Theta^\uparrow_1\Theta^\downarrow_1-\Theta^\uparrow_2\Theta^\downarrow_2),\\
         \left\{\check{R}^{\uparrow\downarrow}_{12}(\mu;\gamma), \Theta^\uparrow_1\Theta^\downarrow_1-\Theta^\uparrow_2\Theta^\downarrow_2\right\}=&2\dc^2(\mu,|\gamma|)\sn(\mu,|\gamma|)\left[h^\uparrow_{12}\Theta^\uparrow_1(\Theta^\downarrow_1-\Theta^\downarrow_2)+h^\downarrow_{12}(\Theta^\uparrow_1-\Theta^\uparrow_2)\Theta^\downarrow_2\right]\\ 
         &+2\gamma  \sn(\mu,|\gamma|) \left[h'^\uparrow_{12}\Theta^\uparrow_1(\Theta^\downarrow_1+\Theta^\downarrow_2)+(\Theta^\uparrow_1+\Theta^\uparrow_2)h'^\downarrow_{12}\Theta^\downarrow_1\right].
    \end{split}
\end{equation}They reduce \eqref{eq:newequation} to four scalar equations
\begin{equation}
    \begin{split}
         \sn(\lambda+\mu)\dc^2(\lambda+\mu)f(\lambda,\mu)\left(1+f(\lambda)f(\mu)\right)&=\sn(\lambda-\mu)\dc^2(\lambda-\mu)\left(f(\lambda)+f(\mu)\right)    ,\\
         \sn(\lambda+\mu)f(\lambda,\mu)\left(1+f(\lambda)f(\mu)\right)&=\sn(\lambda-\mu)\left(f(\lambda)+f(\mu)\right)    ,\\
     \dc(\lambda+\mu)\sn^2(\lambda+\mu)f(\lambda,\mu)\left(1-f(\lambda)f(\mu)\right)&=\dc(\lambda-\mu)\sn^2(\lambda-\mu)\left(f(\lambda)-f(\mu)\right),\\
        \dc(\lambda+\mu)f(\lambda,\mu)\left(1-f(\lambda)f(\mu)\right)&=\dc(\lambda-\mu)\left(f(\lambda)-f(\mu)\right).\label{eq:neweqoff}
    \end{split}
\end{equation}These equations cannot be simultaneously satisfied for all values of $\lambda,\mu$. The reason is that \eqref{eq:newequation} involves four independent operators, the coefficients of each of which must vanish for its LHS and RHS to be identical for arbitrary spectral parameters. Therefore the decorated YBE approach fails to construct an $R$-operator for the superconducting Hubbard Hamiltonian. However, this does not categorically mean that the Hamiltonian \eqref{eq:superH} is necessarily non-integrable. The integrability of this Hamiltonian may be detected from level-spacing statistics \cite{He_2026}, or more rigorously by showing the absence of local conserved charges \cite{PhysRevLett.131.256704}.

\subsection{Compatibility conditions within Shastry’s construction}
Comparison of the preceding examples shows that a further necessary compatibility condition for the superposition ansatz is
\begin{equation}
   \{\check{R}_{12}(\mu),\Theta_1-\Theta_2\}=g_+(\mu) O^{(+)}_{12}, \quad \text{and} \quad  [\check{R}_{12}(\mu),\Theta_1+\Theta_2]=g_-(\mu) O^{(-)}_{12}, \label{eq:condition1}
\end{equation}where $g_\pm(\mu)$ are two functions of the spectral parameter, and $O^{(\pm)}_{12}$ are two independent operators.\footnote{An exception to the discussions that follow is when one of the two functions vanishes. In that case, the superposition ratio $f(\lambda,\mu)$ and hence the $R$-operator cannot be uniquely determined from the Hamiltonian.} This is where the $R$-operator construction for superconducting Hubbard generalization breaks down, as \eqref{eq:condition1} can only be satisfied if $\gamma=0$. Notice that the TRS of the one-parameter $R$-operator implies that $g_+(\mu)$ is even, while $g_-(\mu)$ is odd. Alternatively, the condition can be phrased as
\begin{equation}
\begin{split}
    g_+(\lambda)[\check{R}_{12}(\mu),\Theta_1+\Theta_2]=&g_+(\mu) [\check{R}_{12}(\lambda),\Theta_1+\Theta_2],\\
    \quad g_-(\lambda)\{\check{R}_{12}(\mu),\Theta_1-\Theta_2\}=&g_-(\mu)\{\check{R}_{12}(\lambda),\Theta_1-\Theta_2\}, \quad \forall \lambda,\mu.
\end{split}\label{eq:condition2}
\end{equation}The $R$-operator that appears in the above is that of the free fermion, either of a single copy, as in the case of the XYh model, or the product of the $R$-operators for multiple free fermions, which is the case of the Hubbard model. And the TRO is chosen accordingly.

These additional conditions are not quite enough. Suppose that they are satisfied, then as has been shown multiple times before, the YBE for the $R$-operator of interacting fermions implies the equation on the superposition coefficient
\begin{equation}
    \begin{split}
         g_-(\lambda+\mu)f(\lambda,\mu)\left(1+f(\lambda)f(\mu)\right)&=g_-(\lambda-\mu)\left(f(\lambda)+f(\mu)\right)    ,\\
     g_+(\lambda+\mu)f(\lambda,\mu)\left(1-f(\lambda)f(\mu)\right)&=g_+(\lambda-\mu)\left(f(\lambda)-f(\mu)\right),
    \end{split}\label{eq:condition3}
\end{equation}the ratio between which gives
\begin{equation}
    \frac{f(\lambda)}{1-(f(\lambda))^2}g_1(\lambda,\mu)=\frac{f(\mu)}{1-(f(\mu))^2}g_2(\lambda,\mu), \label{eq:generalf}
\end{equation}where 
\begin{equation}
    \begin{split}
        g_1(\lambda,\mu)=&g_-(\lambda+\mu)g_+(\lambda-\mu)-g_+(\lambda+\mu)g_-(\lambda-\mu),\\
        g_2(\lambda,\mu)=&g_-(\lambda+\mu)g_+(\lambda-\mu)+g_+(\lambda+\mu)g_-(\lambda-\mu).
    \end{split}
\end{equation}For Eq.~\eqref{eq:generalf} to determine the function $f(\lambda)=f(\lambda,0)$, it is necessary that $g_1(\lambda,\mu)/g_2(\lambda,\mu)=G(\mu)/G(\lambda)$ for some function $G(\lambda)$, on a domain where the relevant denominators are nonzero. When this separability condition holds, Eq.~\eqref{eq:generalf} reduces to a one-variable functional equation for $f$, after which $f(\lambda,\mu)$ can be reconstructed from Eq.~\eqref{eq:condition3}. Eqs.~\eqref{eq:condition1}--\eqref{eq:generalf} therefore provide compatibility and separability tests for the restricted YBE, equivalently the $RLL$ relation, within the proposed superposition ansatz. They do not by themselves establish the full three-parameter YBE. The latter requires an independent verification, such as the TZA relations available for the Hubbard and XYh models.

Despite the somewhat involved statement of the final condition, it remains a practical diagnostic for identifying candidate onsite deformations compatible with the proposed superposition ansatz, once the free-fermionic $R$-operator and the decorating involution are known, which are the only inputs to \eqref{eq:condition1} and \eqref{eq:condition2}. If these conditions are satisfied, the superposition coefficient can be determined directly from Eqs.~\eqref{eq:condition1}--\eqref{eq:condition3}, without repeating the preceding model-specific derivations. Combined with the condition and iterative procedure for finding free fermionic $R$-operators, they provide a self-contained procedure for constructing candidate two-parameter $R$-operators at the $RLL$ level, which must subsequently be tested against the full YBE.

\section{Conclusion}\label{sec:conclusion}

In this work, I developed a formal bootstrap and a sequence of compatibility tests for a class of fermionic $R$-operators associated with ordinary and decorated Yang-Baxter equations. The simultaneous expansion of the two equations produces stronger local constraints than expansion of the ordinary YBE alone and recursively determines candidate Taylor coefficients. The essential additional requirement is locality: Expressions generated naturally on three neighboring sites must reduce to operators supported on two sites. This yields a hierarchy of necessary locality conditions for any analytic $R$-operator with the assumed decorated-YBE structure.

I then studied onsite deformations within a Shastry-type superposition ansatz. The ordinary and decorated YBEs reduce the restricted YBE associated with the superposition ansatz to scalar functional equations for the superposition coefficient, which can be solved when an additional separability condition holds. These steps construct a candidate two-parameter $R$-operator satisfying the associated $RLL$ relation within the ansatz. Satisfaction of the full three-parameter YBE remains an additional requirement and, in the Hubbard and XYh examples, follows from established tetrahedral Zamolodchikov algebra relations.

The superconducting Hubbard deformation illustrates the diagnostic value and the limitation of the method. Its operator equation generates four incompatible scalar conditions, proving that the proposed superposition ansatz fails for generic parameters. This does not establish non-integrability of the Hamiltonian under all possible constructions.

It is important to remember that some integrable Hamiltonians that have two-parameter $R$-operators, including those of the XYh and Ising models, also satisfy the Reshetikhin condition. What they all have in common is the additional one-body term in the Hamiltonian in addition to nearest-neighbor interactions. While the current manuscript focuses on fermionic degrees of freedom, Shastry's method has been generalized to SU($n$) Hubbard model by Maassarani \cite{SUNHubbard,fusion}. However, that particular Hamiltonian still resembles $n-1$ copies of fermionic degrees of freedom sharing the same vacuum state. A more interesting generalization may perhaps involve Fock parafermions \cite{PhysRevA.89.012328,zhang2026yangbaxterequationchiralpotts}. Another integrable Hamiltonian with two-parameter $R$-operator is Martins' spin-1 chain \cite{MARTINS2013243}. It would be worth investigating whether that $R$-operator can be constructed directly from the Hamiltonian.

Another natural extension is to introduce anisotropy or superconducting deformation to Maassarani's SU($n$) Hubbard model \cite{SUNHubbard,fusion}, which might result in partial integrability \cite{PhysRevB.106.134420}. One could also consider possible bosonic versions of the models studied in this framework. For instance, the unidirectional hopping Bose-Hubbard model \cite{PhysRevLett.132.086502} has recently been shown to be integrable, and its non-Hermitian Hamiltonian can be considered the effective Hamiltonian of a dissipative Lindblad system \cite{PhysRevResearch.6.L032067}. 

The non-Hermiticity of the integrable deformation of the Hubbard model \eqref{eq:nonHHubbard} obtained from its $R$-operator also lacks more physical understanding. It may be a consequence of the asymmetric distribution of the onsite term in the local Hamiltonian. Meanwhile it is not clear how the decorated-YBE construction might be modified to a Hamiltonian density that is symmetric between two neighboring sites. Another unresolved mystery is that according to the bootstrap program based on the Reshetikhin condition and its higher-order counterparts, the $R$-operator for the XYh model could be of difference form. In the Ising special case, the $R$-operator describes an edge-model embedding in the vertex framework, and it can be alternatively constructed from the Boltzmann weights of the two-dimensional Ising model \cite{MartinsRmat,zhang2026yangbaxterequationchiralpotts}. It is unclear at the moment whether the $R$-operators obtained from the two different methods are equivalent by a smart reparameterization, or there may be more fundamental gaps in the current understanding.

\section*{Author contributions}
The author conceived the project, performed the analytical derivations, and wrote the manuscript. GPT-5.6 Sol has been used to help inspect algebraic derivations in Sec.~\ref{sec:Hubbard} and \ref{sec:XYh}. The author independently reviewed the resulting calculations and remains fully responsible for all claims and conclusions.

\begin{acknowledgements}
I thank Christopher Ekman, Fabian Essler, and Jon Links for valuable discussions. 
\end{acknowledgements}

\appendix

\section{Derivatives of the decorated YBE}\label{sec:sufficiency}

Differentiating the decorated YBE \eqref{eq:DYBE}, one gets the decorated YBE counterpart of the Sutherland equation
\begin{equation}
    \{h_{23},R_{13}(\mu)R_{12}(-\mu)\}=d_\mu R_{13}(\mu)R_{12}(-\mu)-R_{13}(\mu)d_\mu R_{12}(-\mu),\label{eq:aSutherland}
\end{equation}differentiating which leads to the decorated YBE equivalent for the Reshetikhin condition \eqref{eq:dReshetikhin}. The decorated YBE \eqref{eq:DYBE} implies that the operator
\begin{equation}
    \tilde{T}(\mu)=\prod_{j} R_{0,2j}(\mu)R_{0,2j-1}(-\mu)
\end{equation}has the property
\begin{equation}
    \tilde{T}(\mu)\tilde{T}(\nu)=\tilde{T}(-\nu)\tilde{T}(-\mu).\label{eq:anticom} 
\end{equation}Expanding $\tilde{T}(\mu)$ as 
\begin{equation}
    \tilde{T}(\mu)=\sum_{k=0}^\infty \frac{\mu^k}{k!}\tilde{C}_k,
\end{equation}where $\tilde{C}_0=P$ is the translation operator, and $\tilde{C}_1=\sum_{j} (-1)^j h_{j,j+1}$, then \eqref{eq:anticom} implies that 
\begin{equation}
    \tilde{C}_m\tilde{C}_n=(-1)^{m+n}\tilde{C}_n\tilde{C}_m.
\end{equation}That is, $[\tilde{C}_{2k},\tilde{C}_{2l}]=[\tilde{C}_{2k-1},\tilde{C}_{2l-1}]=\{\tilde{C}_{2k},\tilde{C}_{2l-1}\}=0$. Then \eqref{eq:aSutherland} implies that the operator $\tilde{B}=\sum_{j} (-1)^jjh_{j,j+1}$ satisfies
\begin{equation}
    \{\tilde{B}, \tilde{T}(\mu)\}=d_\mu \tilde{T}(\mu),
\end{equation}which in turn implies that $\{\tilde{B},\tilde{C}_k\}=\tilde{C}_{k+1}$.

\section{Anti-Hermitian integrable free fermions}\label{sec:NonHermitian}

A gauge transformation in the bases of Pauli matrices of the XX fermion model reveals non-Hermitian integrable free fermions. Define the rotated Pauli operators by
\begin{equation}
    \begin{split}
        X'_j&=\cos(j\delta)X_j+\sin(j\delta)Y_j,\\
        Y'_j&=-\sin(j\delta)X_j+\cos(j\delta)Y_j.
    \end{split}
\end{equation}The original spin Hamiltonian in the new bases $H(\delta)=\sum_j(X'_jX'_{j+1}+Y'_jY'_{j+1})$ becomes
\begin{equation}
    H(\delta)=\sum_j\left[\cos\delta(X_jX_{j+1}+Y_jY_{j+1})+\sin\delta(X_jY_{j+1}-Y_jX_{j+1})\right].
\end{equation}The JW transformation
\begin{equation}
    X_j= e^{i\pi \sum_{k<j}c^\dagger_kc_k}(c^\dagger_j+c_j),\quad Y_j=-ie^{i\pi \sum_{k<j}c^\dagger_kc_k}(c^\dagger_j-c_j),
\end{equation}brings the Hamiltonian to the fermionic form $H(\delta)=2\sum_jh_{j,j+1}(\delta)$, with
\begin{equation}
    h_{j,j+1}(\delta)=\cos\delta (c^\dagger_jc_{j+1}+c^\dagger_{j+1}c_j)+i\sin\delta(c^\dagger_jc_{j+1}-c^\dagger_{j+1}c_j).
\end{equation}One can easily verify that not only does the (Hermitian) local Hamiltonian $h_{j,j+1}(\delta)$ satisfy the free fermion algebra \eqref{eq:ffalgebra}, but the anti-Hermitian part $h_{j,j+1}=c^\dagger_jc_{j+1}-c^\dagger_{j+1}c_j$ also satisfies the modified free fermion algebra
\begin{equation}
    \{h^2_{12},h_{23}\}=-h_{23}, \quad \{h^2_{23},h_{12}\}=-h_{12}, \quad h^3_{12}=-h_{12}, \quad h_{12}h_{23}h_{12}=0.
\end{equation}The non-Hermitian free fermion Hamiltonian is related to the Hermitian one by conjugation $c^\dagger_jc_{j+1}-c^\dagger_{j+1}c_j=\Theta_j(c^\dagger_jc_{j+1}+c^\dagger_{j+1}c_j)$.

By \eqref{eq:higherR} and the same induction as in Sec.~\ref{sec:test}, its $R$-operator can be iteratively found to be
\begin{equation}
     \check{R}_{12}(\lambda)=1-(\sech(\lambda)-1)h^2_{12}+\tanh(\lambda) h_{12}, \label{eq:RmatNonfreefermion}
\end{equation}which also obeys regularity $\check{R}_{12}(0)=1$ and unitarity $\check{R}_{12}(\lambda)\check{R}_{12}(-\lambda)=1$. In addition, since all three terms commute with one another, it composes as
\begin{equation}
    \check{R}_{12}(\lambda)\check{R}_{12}(\mu)=\check{R}_{12}(\mu)\check{R}_{12}(\lambda)=\check{R}_{12}(\arctanh(\tanh\lambda\sech\mu+\tanh\mu\sech\lambda)),\label{eq:Noncomposition}
\end{equation}which by the same duality transformation \eqref{eq:dualtransform} can be expressed as 
\begin{equation}
    \check{R}^*_{12}(\lambda^*)\check{R}^*_{12}(\mu^*)=\check{R}^*_{12}(\mu^*)\check{R}^*_{12}(\lambda^*)=\check{R}^*_{12}(\lambda^*+\mu^*).
\end{equation}This new $R$-operator results in a separate set of conserved charges, and that is why the XX free fermion has two independent mastersymmetries and consequently two infinite series of local conserved charges \cite{sunXX}.

\bibliographystyle{quantum}
\bibliography{DYBE.bib}
\end{document}